\documentclass[11pt]{article}
\usepackage[a4paper]{geometry}
\usepackage{xcolor}
\usepackage{pstricks}
\usepackage{ctable}
\usepackage{multirow}
\usepackage{amsmath}
\usepackage{hyperref}
\usepackage{dsfont}

\makeatletter
\def\section{\@startsection{section}{1}{\z@}{1ex}{1ex}{\bf \large}}
\makeatother

\newfont{\bbbold}{msbm10}

\def\bbC{\mbox{\bbbold C}}

\def\bbR{\mbox{\bbbold R}}

\def\cE{{\cal E}}

\def\cN{{\cal N}}

\def\cV{{\cal V}}

\newfont{\goth}{eufm10 scaled \magstep1}

\def\ge{\mbox{\goth e}}

\def\gg{\mbox{\goth g}}

\def\gl{\mbox{\goth l}}

\def\gs{\mbox{\goth s}}

\def\gu{\mbox{\goth u}}

\def\a{\alpha}\def\adt{\dot \alpha}
\def\b{\beta}\def\bdt{\dot \beta}
\def\c{\gamma}\def\cdt{\dot\gamma}
\def\d{\delta}
\def\ve{\varepsilon}
\def\F{\Phi}

\def\s{\sigma}

\def\th{\theta}

\def\O{\Omega}
\def\o{\omega}
\def\ua{{\underline{\alpha}}}
\def\ub{{\underline{\phantom{\alpha}}\!\!\!\beta}}
\def\uc{{\underline{\phantom{\alpha}}\!\!\!\gamma}}

\def\ud{{\underline{\delta}}}

\def\xz{\times}

\def\tE{\tilde{E}}

\def\bsh{\backslash}
\def\ie{{\it i.e.}\ }

\def\tE{\widetilde{E}}

\def\nn{\nonumber}
\def\bd{\begin{document}}
\def\ed{\end{document}}
\def\be{\begin{equation}}
\def\ee{\end{equation}}
\def\ba{\begin{array}}
\def\ea{\end{array}}
\def\bea{\begin{eqnarray}}
\def\eea{\end{eqnarray}}
\def\ft#1#2{\tfrac{#1}{#2}}
\def\fft#1#2{\frac{#1}{#2}}
\def\sst#1{{\scriptscriptstyle #1}}
\def\oneone{\rlap 1\mkern4mu{\rm l}}

\newcommand{\eq}[1]{(\ref{#1})}
\newcommand{\w}[1]{\\[0.#1cm]}
\def\eqs#1#2{(\ref{#1}-\ref{#2})}
\def\det{{\rm det\,}}
\def\tr{{\rm tr}}
\def\del{\partial}
\def\demi{{\scriptscriptstyle 1/2}}
\def\quart{{\scriptscriptstyle 1/4}}
\newcommand{\ord}[1]{{\scriptscriptstyle (#1)}}

\newcommand{\hoch}[1]{$\, ^{#1}$}
\newcommand{\imperial}{\it\small Theoretical Physics Group, Imperial College London\\ Prince Consort Road, London SW7 2AZ, UK}
\newcommand{\kings}
{\it\small Department of Mathematics, King's College, University of London\\ Strand, London WC2R 2LS, UK}
\newcommand{\uu}
{\it\small Department of Theoretical Physics, Uppsala, Sweden}
\newcommand{\golm}
{\it\small AEI, Max Planck Institut f\"ur Gravitationsphysik\\ Am M\"{u}hlenberg 1, D-14476 Potsdam, Germany}
\makeatletter
\newcommand{\lapp}
{\it\small LAPTH, Universit{'e} de Savoie, CNRS, B.P. 110, F-74941 Annecy-le-Vieux Cedex France}
\newcommand{\dur}
{\it\small Institute for Particle Physics Phenomenology, Department of Mathematical Sciences and Department of Physics, Durham University, Durham DH1 3LE UK}
\renewcommand\theequation{\thesection.\arabic{equation}}
\@addtoreset{equation}{section} \makeatother

\newcommand{\auth}{\large G. Bossard\hoch{1}, P.S.\ Howe\hoch{2} and K.S. Stelle\hoch{3}}
\begin{document}

\renewcommand{\thefootnote}{\fnsymbol{footnote}}

\null
\begin{flushright}
{\small AEI-2010-143}\\
{\small KCL-MTH-10-09}\\
{\small Imperial/TP/10/KSS/02}\\
\vskip 1.5 cm
\end{flushright}

\begin{center}
{\Large{\bf  On duality symmetries of supergravity invariants }}
\vspace{.75cm}

\auth

\vspace{.5cm}

\begin{itemize}
\item [$^1$]\golm\item[$^2$] \kings \item [$^3$] \imperial
\end{itemize}
\vspace{1cm}

{\bf Abstract}
\end{center}

The role of duality symmetries in the construction of counterterms
for maximal supergravity theories is discussed in a field-theoretic
context from different points of view. These are: dimensional
reduction, the question of whether appropriate superspace measures
exist and information about non-linear invariants that can be gleaned
from linearised ones. The former allows us to prove that F-term
counterterms cannot be $E_{7(7)}$-invariant in $D=4, N=8$
supergravity or $E_{6(6)}$-invariant in $D=5$ maximal 
supergravity. This is confirmed by the two other methods which
can also be applied to $D=4$ theories with fewer supersymmetries and allow us to prove
that $N=6$ supergravity is finite at three and four loops and that
$N=5$ supergravity is three-loop finite.

\vskip .5cm
\vspace{1cm}

\hrule

\null

\renewcommand{\thefootnote}{\arabic{footnote}}
\setcounter{footnote}{0}

\pagebreak
\setcounter{page}{1}

\section{Introduction}

It has now been established that $D=4, N=8$ supergravity is finite at three loops \cite{Bern:2007hh}, despite the existence of a linearised $R^4$ counterterm \cite{Kallosh:1980fi,Howe:1981xy} generalising the $(\hbox{Bel-Robinson})^2$ type counterterm of the $N=1$ theory \cite{Deser:1977nt}. It is also now known that maximal supergravity is finite at four loops in $D=5$ \cite{Bern:2009kd}. The only other candidate linearised short counterterms (\ie F  or BPS terms)
in $D=4$ occur at the five and six loop orders and are four-point terms of the form $\partial^{2k} R^4$ for $k=2,3$ \cite{Drummond:2003ex,Elvang:2010jv,Drummond:2010fp}. The absence of the $R^4$ divergence can be seen from field-theoretic arguments \cite{Howe:2002ui,Bossard:2009sy}, including algebraic renormalisation theory, results that generalise those for the finiteness of one-half BPS counterterms in maximal super Yang-Mills theories in various dimensions \cite{Bossard:2009sy}. However, even in the Yang-Mills case it does not seem easy to extend these results to the double-trace $\partial^2 F^4$ invariant \cite{Bossard:2009mn,bhlsw} which is known to be finite at three loops in $D=6$ \cite{Bern:2010tq}. String theory provides an alternative approach to discussing field-theoretic finiteness issues and has been used to give arguments in favour of the known Yang-Mills results and also suggesting that $D=4, N=8$ supergravity should be finite at least up to six loops \cite{Green:2010sp}. In \cite{Bjornsson:2010wm} a similar conclusion was reached using a first-quantised approach to supergravity based on pure spinors.

A key feature of supergravity theories which has no analogue in SYM is the existence of duality symmetries. It has recently been shown \cite{Bossard:2010dq} that $E_{7(7)}$ can be maintained in perturbation theory in $D=4$ (at the cost of manifest Lorentz invariance), and this suggests that these duality symmetries should be taken seriously in providing additional constraints on possible counterterms which might not be visible from a linearised analysis. For $R^4$, a scattering amplitude analysis supporting the idea that the full invariant is not compatible with $E_{7(7)}$ was given in \cite{Brodel:2009hu}, while in a recent paper this violation of  $E_{7(7)}$ invariance was demonstrated by means of an argument based on dimensional reduction from type II string theory \cite{Elvang:2010kc}.

In this article, we investigate this issue for $D=4, N=8$ supergravity in a field-theory setting from three different points of view: dimensional reduction of higher-dimensional counterterms, the (non)-existence of appropriate superspace measures to generalise the linearised ones, and use of the so-called ectoplasm formalism which allows one to write super-invariants in terms of closed superforms. This analysis implies in particular that the requirement of $E_{7(7)}$ invariance postpones the onset of UV divergences in the $D=4$, $N=8$ theory until at least seven loops. Similar analysis of the $D=4$, $N=5$ and $N=6$ theories shows that they will be finite through three and four loops respectively.

\section{Dimensional reduction}

One way of stating the problem with duality invariance is to start from the $R^4$ counterterm in $D=11$ and to reduce it to $D=4$. This reduced invariant will only have the natural $SO(7)$ symmetry of a standard Kaluza-Klein reduction on $T^7$. However, the invariant may be promoted to a full $SU(8)$ invariant by first performing the necessary dualisations of higher-form fields and then averaging, \ie parametrising the embedding of $SO(7)$ into $SU(8)$ and integrating over the $SU(8)/SO(7)$ coset in a fashion similar to that employed in harmonic superspace constructions. If the Kaluza-Klein reduction ansatz (using the notation of Ref.\ \cite{Lu:1995yn}) for the metric in a $T^{11-D}$ reduction from 11 to $D$ dimensions takes the form  $ds_{\scriptscriptstyle 11}^{\; 2}=e^{\frac13\vec g\cdot\vec\phi}ds_{\scriptscriptstyle D}^{\; 2}+\ldots$\;, the $T^{11-D}$ compactification volume is proportional to $e^{-\frac{(D-2)}6\vec g\cdot\vec\phi}$. For the Einstein action, the $T^{11-D}$ volume factor $e^{-\frac{(D-2)}6\vec g\cdot\vec\phi}$ cancels against a factor $e^{\frac16 D\vec g\cdot\vec\phi}e^{-2
\frac13\vec g\cdot\vec\phi}$ arising out of $\sqrt{-g_\sst D}g_{\sst D}^{\mu\nu}$. However, for the $R^4$ invariant in $D=11$, the Einstein-frame reduction produces an extra factor of $e^{-\vec g\cdot\vec\phi}$ arising with the three additional inverse metrics present in the $R^4$ term as compared to the Einstein-Hilbert action.  This dilatonic factor will then be promoted to an $SU(8)$ invariant by the $SU(8)/SO(7)$ integration. If we expand the exponentials in power series, terms linear in the scalars vanish in such an averaging, but $SU(8)$ invariant quadratic terms survive.

Let us illustrate this in the simpler case of the dimensional reduction of the 11-dimensional $R^4$ invariant down to 8 dimensions, where the duality group is $SL(3,\bbR)\times SL(2,\bbR)$ while the linearly realised subgroup is $SO(3)\times SO(2)$. The dimensionally reduced invariant is manifestly $SL(3,\bbR)$ invariant by construction, but the resulting gravitational $R^4$ term in $D=8$ is multiplied by the $SO(2)$ non-invariant prefactor $e^{-v}$ (we define the volume modulus $v \equiv \frac1{||g||}\vec g\cdot\vec\phi$ for convenience), while the Pontryagin term  $\frac{1}{4} p_1^2 - p_2$ is multiplied by the $SO(2)$ non-invariant axion field $a$ descending from the $3$-form in $D=11$. Integrating the  $SO(2)$ transformed complex function $\tau \equiv a + i e^{-v} $ over $SO(2)$, one obtains 

\be  
\int_{- \frac{\pi}{2}}^{\frac{\pi}{2}}  \frac{d \theta}{\pi}   \, \frac{ \tau - \tan \theta }{1 + \tau \tan \theta }Ê= \int_{-\infty}^{+\infty} \frac{dt}{\pi} \, \frac{ \tau}{1 - t^2 \tau^2 }Ê= i \; \; .  \label{D8invariant}
\ee

One therefore concludes that the $SO(2)$ invariant $R^4$ terms do not in fact depend on the scalar fields; in particular, the prefactor of the parity-even $R^4$ term averages to a constant and the coefficient of the Pontryagin term averages to zero. This is as expected because this invariant is associated to the known 1-loop logarithmic divergence of maximal supergravity in eight dimensions, and it must therefore preserve $SL(2,\bbR)$ invariance.

However, we are going to see that the eight-dimensional case is rather special, and that one generally obtains a non-trivial function of the scalar fields after reducing to lower dimensions and R-symmetry averaging. The first non-trivial example appears in seven dimensions. The scalar fields are the volume modulus $v$ and the $SL(4,\bbR)$ symmetric matrix $G_{IJ}$ which together constitute the moduli of $T^4$, plus four axion fields $a^I$ descending from the 11-dimensional 3-form. The $R^4$  term resulting from the eleven dimensional invariant after dimensional reduction is multiplied by $ e^{-||g||v}=e^{-4\zeta}$, where we have defined $\zeta \equiv  \frac14Ê\vec g\cdot\vec\phi=\frac14 ||g||v$. One computes the $SO(5)$ averaging integral (in terms of stereographic coordinates $t^I$ for $S^4 \cong SO(5) / SO(4)$)

\bea & &Ê\frac{3}{4\pi^2}Ê\int \frac{ d^4 t}{ ( 1 + t^2 )^\frac{5}{2} } Ê \, \frac{Ê( 1 + t^2 ) \, e^{-4 \zeta}Ê}{\bigl(  1 + t^I a_I \bigr) ^2  + e^{-5\zeta} G_{IJ} t^I t^J } \nn\\
 &=& 1 -  \frac{6}{7}Ê\zeta^2 - \frac{3}{35} a^I a_I + \dots   \nn\\
&=& 1 - \frac{6}{35}Ê\phi_{AB}Ê\phi^{AB}Ê+ \mathcal{O}(\phi^4) \; \; , 
\eea

where $\phi_{AB}$ parametrise $SL(5,\bbR) / SO(5)$ in the symmetric gauge.\footnote{More specifically 
\be  \mbox{exp}[Ê2 \phi_{AB} ]Ê = \left( \begin{array}{cc} \; e^{4\zeta}Ê\; & \; e^{4\zeta} a_J \; \vspace{1mm} \\ \; e^{4\zeta} a_I  \; & \; e^{-\zeta}ÊG_{IJ} + e^{4\zeta}Êa_I a_J \; \end{array}\right) \ . \nn \ee} This function clearly depends on the scalar fields, and is thus not invariant with respect to $SL(5,{\bbR})$.

 In lower dimensions, similar explicit evaluations of the scalar prefactors are possible in principle, but burdensome in practice. Thus, we will adopt a different approach, which at the same time clarifies the relation to the string-theory based discussions of Refs \cite{Green:2010sp,Elvang:2010kc}. To set the stage, we return first to $D=10$ and consider the scalar prefactor appearing there for the $R^4$ type invariant. In $D=10$, one finds $g=\frac12$, corresponding to a $D=11\rightarrow D=10$ reduction with an $S^1$ circumference proportional to $e^{-\frac23 \phi}$, where $\phi$ is the type IIA dilaton. The dimensional reduction of the $D=11$ $R^4$ invariant generates a $D=10$ scalar prefactor $e^{-\frac12\phi}$, so one sees immediately that this $D=10$ supersymmetry invariant cannot also be invariant under the continuous $GL(1)$ ``duality'' symmetry of the type IIA theory. Since the $D=10$ $R^4$ invariant is not called for as a counterterm by na\"{\i}ve power counting, this does not have a direct bearing on type IIA supergravity infinities. However, what interests us here is the pattern of $GL(1)$ non-invariance. In the $D=10$ case, one may capture a relevant feature of the $e^{-\frac12\phi}$ scalar prefactor by noting that on the type IIA $GL(1)/\{ \oneone\} $ scalar target manifold, it satisfies the Laplace equation

\be
\left(\frac{ \partial^2 \, }{\partial\phi^2 }-\frac14 \right)e^{-\frac12\phi}=0\ .
\ee

It is clear from the nonzero Laplace eigenvalue in this equation that a $GL(1)$ invariant function $f(\phi)=1$ cannot be a solution to this equation, so the continuous $GL(1)$ duality symmetry is necessarily broken by this $D=10$ counterterm.

 In $D=9$, maximal supergravity has a three-dimensional space of scalars taking their values in a $GL(2,\bbR)/SO(2)$ target space, with a scalar sector Lagrangian 

\be L_{\rm scalar}=\sqrt{-g}(-\frac12\partial_\mu\phi_1\partial^\mu\phi_1-\frac12\partial_\mu\phi_2\partial^\mu\phi_2-\frac12e^{-\frac32\phi_1+\frac{\sqrt7}2\phi_2}\partial_\mu\chi\partial^\mu\chi)\ee

which may be rewritten as 

\be
L_{\rm scalar}=\sqrt{-g}(-\frac12\partial_\mu v_2\partial^\mu v_2 - \frac12\partial_\mu\varphi\partial^\mu\varphi - \frac12e^{2\varphi}\partial_\mu\chi\partial^\mu\chi)\ ,\ee

 where the $T^2$ volume modulus is $v_2=\frac1{||g||}\vec g\cdot\vec\phi=\frac{\sqrt7}4\phi_1+\frac34\phi_2$ (in terms of which the $T^2$ reduction volume is proportional to $e^{-\frac76\vec g\cdot\vec\phi}=e^{-\frac{\sqrt7}3v_2}$) and the orthogonal dilatonic combination is $\varphi=-\frac34\phi_1+\frac{\sqrt7}4\phi_2$, while $\chi$ is the $D=9$ axionic scalar emerging from the $D=10$ Kaluza-Klein vector. The $GL(2,\bbR)$ invariant Laplace operator on the $GL(2,\bbR)/SO(2)$ scalar target space thus becomes
 
\be
\Delta_9=\frac{\partial^2 \,}{\partial v_2^2 }+\left(\frac{\partial^2 \, }{\partial \varphi^2 } + \frac\partial{\partial\varphi}+ e^{-2\varphi}\frac{\partial^2 \, }{\partial \chi^2 }\right)\ .\label{d9laplace}
\ee

The bracketed terms in \eqref{d9laplace} may be recognised as the Laplace operator on $SL(2,\bbR)/SO(2)$. The scalar prefactor of the $R^4$ term descending from $D=11$ in this case is $e^{-3(\phi_1+\frac3{\sqrt7}\phi_2)}=e^{-\frac2{\sqrt7}v_2}=e^{-||g||v_2}$. This scalar prefactor does not depend on the $(\varphi,\chi)$ fields of the $SL(2,\bbR)/SO(2)$ sector, however, so the $SL(2,\bbR)/SO(2)$ Laplace operator does not contribute when acting upon this function. The Laplace equation satisfied by the $R^4$ scalar prefactor in $D=9$ is thus

\be
(\Delta_9 -||g||_9^2)e^{-||g||_9v_2}=0\ ,\quad ||g||_9=\frac2{\sqrt7}\ .\label{laplace9eqn}
\ee

Comparing to the $D=10$ case with $||g||_{10}=\frac12$ and $v_1=\phi$, we see that the Laplace equation satisfied by the $f(v)$ prefactor has the same form in the $D=10$ and $D=9$ cases. 
The $D=9$ equation \eqref{laplace9eqn} is invariant under the $H_9=SO(2)$ R-symmetry of $D=9$ supergravity .  The $H_9=SO(2)$ R-symmetry also leaves the $v_2$ volume modulus invariant in this case. It is clear, however, from the $||g||^2$ ``mass term'' in \eqref{laplace9eqn} that a $GL(2,\bbR)$ invariant (\ie constant) function $f(v_2)=1$ cannot be a solution of this equation, so the continuous $D=9$ $GL(2,\bbR)$ duality symmetry is necessarily broken by the $R^4$ counterterm in $D=9$.

When one descends one step further to $D=8$, a rather special case arises. Once again, the scalar prefactor for the $R^4$ term derived directly from a $T^3$ $D=11\rightarrow D=8$ reduction depends only on the volume modulus $v_3=\frac1{||g||}\vec g\cdot\vec\phi=\frac12\phi_1+\frac3{2\sqrt7}\phi_2+\sqrt{\frac37}\phi_3$. The target manifold for seven scalar fields in $D=8$ has structure $(SL(3,\bbR)\times SL(2,\bbR))/(SO(3)\times SO(2))$. As in $D=10\;\&\;9$, the $v_3$ volume modulus is clearly invariant under the ``gravity line'' $SO(3)$ little group. However, the total set of seven $D=8$ scalars is now completed for the first time by an extra axionic scalar $a=A_{11\,10\,9}$ emerging from the $D=11$ 3-form gauge field. This axionic scalar together with the $T^3$ volume modulus $v_3$ parametrise the $SL(2,\bbR)/SO(2)$ portion of the scalar-field target space. 
Defining the orthogonal dilatonic combinations $\varphi_1=-\frac34\phi_1+\frac{\sqrt7}4\phi_2$ and $\varphi_2=-\frac{\sqrt3}2\phi_1-\frac{3\sqrt3}{4\sqrt7}\phi_2+\frac2{\sqrt7}\phi_3$ and letting $\chi_1$, $\chi_2$ and $\chi_3$ be the axions emerging from Kaluza-Klein vectors, one finds the $SL(3,\bbR)\times SL(2,\bbR)$ invariant $D=8$ Laplace operator

\bea
\Delta_8&=&\left(\frac{\partial^2}{\partial v_3^2}+\frac{\partial}{\partial v_3}+e^{-2v_3}\frac{\partial^2}{\partial a^2}\right)\nn\\
&&\hspace{-.3cm}+\left(\frac{\partial^2}{\partial\varphi_1^2}+\frac{\partial}{\partial\varphi_1}+\frac{\partial^2}{\partial\varphi_2^2}+\sqrt3\frac{\partial}{\partial\varphi_2}+e^{-2\varphi_1}\frac{\partial^2}{\partial\chi_1^2}+e^{-\varphi_1-\sqrt3\varphi_2}\frac{\partial^2}{\partial\chi_2^2}+e^{\varphi_1-\sqrt3\varphi_2}\frac{\partial^2}{\partial\chi_3^2}\right)\ .\label{d8laplace}
\eea

The first line in Eq.\ \eqref{d8laplace} comprises the Laplace operator on $SL(2,\bbR)/SO(2)$ while the second line comprises the Laplace operator on $SL(3,\bbR)/SO(3)$. 

Dimensionally reducing the $R^4$ term from $D=11$ to $D=8$ produces an $R^4$ term with a scalar prefactor $e^{-||g||v_3}$ just as in $D=10$\;\&\;$D=9$. However, the $SL(2,\bbR)/SO(2)$ part of the Laplace operator involving the single field $v_3$ on which this prefactor depends is now $\frac{\partial^2}{\partial v_3^2}+\frac\partial{\partial v_3}$ owing to the nonlinear dependence of the $SL(2,\bbR)/SO(2)$ target space metric on the volume modulus $v_3$. Moreover, in $D=8$ one has $||g||_8=1$. Consequently, the $D=8$ Laplace equation satisfied by the scalar prefactor $f(v_3)$ is
\be
\Delta_8\,e^{-||g||_8v_3}=0\ ,\quad ||g||_8=1\label{laplace8eqn}
\ee
with a vanishing ``mass term''.

As we have already seen, although the $D=8$ prefactor $f(v_3)=e^{-v_3}$ obtained directly by dimensional reduction from $D=11$ is properly $SO(3)$ invariant, it fails to be $SO(2)$ invariant since $SO(2)$ mixes $v_3$ and $a$. However any $SO(2)$ transform of the $D=8$ $R^4$ type invariant would equally well satisfy the requirements of local supersymmetry and gauge invariance, and moveover, since the Laplace equation \eqref{laplace8eqn} is fully invariant under $SO(3)\times SO(2)$, such an $SO(2)$ transform would satisfy Eq.\ \eqref{laplace8eqn} as well. Consequently, averaging over such $SO(2)$ transforms in order to produce a fully $SO(3)\times SO(2)$ invariant candidate counterterm must also give a result satisfying the Laplace equation \eqref{laplace8eqn}. As we have seen in Eq.\ \eqref{D8invariant}, this averaging in fact produces $f(v_3,a)=\hbox{constant}$, which is allowed by \eqref{laplace8eqn}. Thus the pure $R^4$ structure of the averaged $D=8$ candidate counterterm is fully $SL(3,\bbR)\times SL(2,\bbR)$ invariant. This is as it should be, since there is a known $R^4$ divergence at one loop in $D=8$ maximal supergravity.

Now let us continue down to $N=8$ supergravity in four dimensions, where the target space for the 70 scalars is $E_{7(7)}/SU(8)$. The action of $SU(8)$ on the $T^7$ volume modulus field $v=\vec g\cdot\vec\phi$ is highly nonlinear, so we will not present an explicit $SU(8)$ averaging of the dimensionally reduced $R^4$ invariant for the $D=4$ case. Instead, we shall now concentrate on the $E_{7(7)}$ invariant Laplace equation which must be satisfied, corresponding to \eqref{d9laplace} and \eqref{d8laplace} for the $D=9$ and $D=8$ cases. The scalar prefactor of the $R^4$ term obtained via dimensional reduction from $D=11$ on $T^7$ is $e^{-||g||_4v}$, with $||g||_4=\sqrt7$.

After dualisation, the 70 scalar fields parametrise $E_{7(7)} / SU(8)_{\scriptscriptstyle \rm c}$ in a gauge associated to the parabolic subalgebra

\be \bigl(  \gg\gl_1 \oplus \gs\gl_7 \bigr)^\ord{0}\oplus {\bf 35}^\ord{3}\oplus  \overline{\bf 7}^\ord{6}  \subset \ge_{7(7)} \; \; , \ee

together  with any convenient choice of gauge for the $SL(7,\bbR) / SO(7) $ scalars.

By construction, the Laplace equation on the scalar target manifold is invariant with respect to the non-linear action of $SU(8)$, and it follows that the $SU(8)$ invariant scalar-field prefactor\footnote{The subscript on the $D=4$ scalar prefactor indicates the loop order at which the invariant would be expected to occur under na\"{\i}ve power counting.} $f_3$ occurring in the $D=4$ invariant satisfies the same Laplace equation as $e^{-\sqrt7 v}$:

\be
 \Delta f_3(\phi) = \Delta \int_{\scriptscriptstyle SU(8)/SO(7)} \hspace{-12mm} du \hspace{10mm}  e^{- \sqrt7 v(u)} =  \int_{\scriptscriptstyle  SU(8)/SO(7)} \hspace{-12mm} du \hspace{10mm}  \Bigl( \Delta \;  e^{- \sqrt7 v(u)} \Bigr)(u)\; \; .  
\ee

For the dimensionally reduced $R^4$ invariant, one finds that the scalar-field prefactor $f_3$  satisfies the Laplace equation

\be \Delta \, e^{-\sqrt7 v} =   \left(\frac{ \partial^2 \, }{\partial v^2 } + 7\sqrt7Ê \frac{\partial\, }{\partial v }\right) e^{-\sqrt7 v} = - 42 \, e^{-\sqrt7 v} \; ,  \ee

and so 

\be \bigl(  \Delta + 42  \bigr) f_3(v) = 0\; , \label{Laplace3} \ee

in agreement with the computation of \cite{Elvang:2010kc}. It is clear that $f_3(v)=1$ is not a solution of the Laplace equation (\ref{Laplace3}), and so we conclude that the unique $R^4$ invariant is not $E_{7(7)}$ invariant in four dimensions. 

Considering the seven independent $SU(8)$ invariant functions of the 70 linearly-transforming scalar fields $\phi_{ijkl}$, one can easily convince oneself that the only $SU(8)$ invariant solution to the Laplace equation $\Delta f_0(\phi) = 0 $ is $f_0(\phi) = 1$. It follows that the Laplace equation $( \Delta + 42 ) f_3(\phi) = 0$ 
determines uniquely $f_3(\phi)$ as a formal power series in $\phi_{ijkl}$ (note that we are only interested here in the Taylor expansion of $f_3(\phi)$ in perturbation theory).

Dilaton factors of this sort in front of purely gravitational terms constructed from curvatures and their covariant derivatives prevent such terms from being constituent parts of duality invariants, since the lowest-order part of a duality transformation involves constant shifts of the scalar fields. Of course, if there were additional invariants arising in a given spacetime dimension, without Kaluza-Klein origins, combinations of such invariants might be capable of erasing the problematic dilatonic scalar prefactors, thereby permitting a duality-invariant construction. However, in $D=4$, the only available 1/2 BPS $SU(8)$ invariant $R^4$ counterterm \cite{Howe:1981xy} is {\em unique} at the 4-point level \cite{Drummond:2003ex}. This counterterm develops a higher-point structure which is not fully known, but this higher-point structure must also be unique. Were there alternative higher-point structures extending this 4-point linearised supersymmetry invariant, their differences would themselves have to constitute new $D=4$ invariants under $SU(8)$-covariant linearised supersymmetry, and these do not exist \cite{Drummond:2003ex}. Thus, the uniqueness of the $SU(8)$-symmetric $R^4$ invariant in $D=4$ maximal supergravity shows that the $SU(8)$-symmetrised dimensional reduction of the $R^4$ invariant in $D=11$ is the only such supersymmetric candidate. Its ineligibility as an $E_{7(7)}$ duality invariant thus rules out the $R^4$ candidate counterterm in $D=4$.

The above argument is a variant of the one given in Ref.\ \cite{Elvang:2010kc} (where it was framed in terms of reduction from $D=10$ type II superstring/supergravity amplitudes). It also gives a way to see that the maximal supergravity 1/4 BPS $\partial^4R^4$ candidate counterterm at 5 loops and the 1/8 BPS $\partial^6R^4$ candidate counterterm at 6 loops cannot be $E_{7(7)}$ duality invariants either. Once again, the argument hangs upon the uniqueness of the corresponding $D=4$ $SU(8)$ symmetric BPS invariants, together with the Laplace equation for the dilaton factors arising from dimensional reduction in front of the purely gravitational parts of the invariants. 

In fact, it is precisely the known existence of the 1/2 BPS $R^4$ one-loop divergence of maximal supergravity in $D=8$, the 1/4 BPS $\partial^4R^4$ two-loop divergence in $D=7$ and the 1/8 BPS $\partial^6R^4$ three-loop divergence in $D=6$ that permits us to rule out the descendants of these counterterms as $E_{7(7)}$ invariants in $D=4$. The existence of these higher-dimensional divergences indicates the presence of corresponding counterterms without dilaton factors in the purely gravitational parts of the higher-dimensional versions of these counterterms. Indeed, the demonstration that $E_{7(7)}$ symmetry is preserved in perturbative theory for $N=8$ supergravity \cite{Bossard:2010dq}, generalises straightforwardly to higher dimensions, provided that there are no Lorentz $\times$ $R$-symmetry one-loop anomalies.  The absence of such an anomaly is trivial in odd dimensions, and there is none in six dimensions \cite{Marcus}. The $SL(2,\bbR)$ symmetry is, admittedly, anomalous at the one-loop order in $D=8$, but the latter does not affect the consequences of the tree-level Ward identities for the one-loop divergence, and this must therefore be associated to an $SL(2,\bbR)$ duality-invariant $R^4$ counterterm, as we have seen above. Coupled with the $D=4$ uniqueness of the $R^4$, $\partial^4R^4$ and $\partial^6R^4$ BPS counterterms \cite{Drummond:2003ex}, the inevitable appearance of dilaton factors in the $D=4$ versions of these counterterms then rules out $E_{7(7)}$ invariance for all these BPS candidate operators.

To see how this works more generally, define the Kaluza--Klein Ansatz for  reduction from a higher dimension $D=11,\ 8,\ 7,\ 6,\ 5$ down to $D=4$: 

\be ds_{\scriptscriptstyle D}^{\; 2} = e^{(D-4) \phi_{\scriptscriptstyle D}}  ds_{\scriptscriptstyle 4}^{\; 2} + e^{-2 \phi_{\scriptscriptstyle D}} G_{IJ} ( dy^I + A^I ) ( dy^J + A^J ) \; . \label{KKmetric}\ee

Since we will be treating a number of spacetime dimensions at once, it is convenient at this stage to adopt a normalisation for the scalar fields different from the canonical normalisation we have used so far. This will not, however, affect the key values of the Laplace ``masss terms'' that we seek to derive. Each axion field, labelled $l_a$, originating from the dimensional reduction of a form field in $D$ dimensions, admits a kinetic term with a factor $e^{2 w_D(l_a) \phi_{\scriptscriptstyle D}}$, and each vector field, labelled by $l_v$, admits a kinetic term with a factor $e^{2 w_D(l_v) \phi_{\scriptscriptstyle D}}$ as for example, starting from $D=11$,  with $\phi_{11}=\frac{v}{3||g||}=\frac{v}{3\sqrt7}$ in terms of our earlier $R^4$ discussion, 

\begin{multline}Ê e^{6 \phi_{\scriptscriptstyle 11}} G^{IL} G^{JM} G^{KN}Êd a_{IJK}\wedge \star d a_{LMN} + e^{12 \phi_{\scriptscriptstyle 11}} G_{IJ}Êd a^I \wedge\star d a^J \\* +  e^{- 3 \phi_{\scriptscriptstyle 11}} G^{IK} G^{JL}ÊF_{IJ}Ê\wedge\star F_{KL}Ê+ e^{-9 \phi_{\scriptscriptstyle 11}} G_{IJ} F^{I}Ê\wedge\star F^J\ .   \end{multline}

 One then computes how the $SU(8)$ invariant Laplace operator acts on the function $e^{-(D-4) n \phi_{\scriptscriptstyle D}}$ multiplying the purely gravitational $\partial^{2(n-3)} R^4$ terms after dimensional reduction from $D$ to 4 dimensions:
 
\bea \Delta \; e^{-(D-4) n \,  \phi_{\scriptscriptstyle D}} &=& \frac{3}{\sum_{l_v} w_{\scriptscriptstyle D}
(l_v)^2}  \left(  \frac{ \partial^2 \, }{\partial \phi_{\scriptscriptstyle D}{}^2 } +\biggl( \sum_{l_a} w_{\scriptscriptstyle D}(l_a)Ê\biggr)  \frac{\partial\, }{\partial \phi_{\scriptscriptstyle D} }\right) e^{-(D-4) n \,  \phi_{\scriptscriptstyle D}} \nn \\  &=&  \frac{D-4}{D-2} n ( D + n - 32 ) \; e^{-(D-4) n \, \phi_{\scriptscriptstyle D}} \; .  \eea

If it is assumed that there is a duality-invariant $\partial^{2(n-3)} R^4$  type counterterm in $D$ dimensions, then the corresponding  scalar field function $f_n(\phi)$ multiplying $\partial^{2(n-3)} R^4$ in four dimensions has to satisfy 

\be  \left( \Delta+  \frac{D-4}{D-2} n ( 32-  D - n  ) \right) f_n(\phi) = 0 \; . \label{dualitycond}\ee

We have already seen in Eq.\ \eq{Laplace3} how the existence of an $SL(3,\bbR)\times SL(2,\bbR)$ duality-invariant one-loop divergence in $D=8$ maximal supergravity implies $f_3(\phi)\ne 1$ for the unique $SU(8)$-invariant $R^4$ type counterterm in $D=4$, showing that this potential 3-loop 1/2 BPS counterterm cannot be $E_{7(7)}$ invariant. Similarly, the known 2-loop divergence of maximal supergravity in $D=7$, which has $SL(5,\bbR)$ duality invariance, implies that the function $f_5(\phi)$ multiplying $\partial^{4} R^4$ in $D=4$ invariant must satisfy 

\be ( \Delta + 60 ) f_5(\phi) = 0 \; . \ee

This then implies that the unique 1/4 BPS $SU(8)$-invariant $\partial^{4} R^4$ type counterterm in $D=4$ cannot be duality-invariant. And similarly, from the existence of a 3-loop divergence in $D=6$ maximal supergravity with $SO(5,5)$ duality invariance, one learns that the function $f_6(\phi)$ multiplying $\partial^6 R^4$ in the corresponding $D=4$ counterterm must satisfy
\be ( \Delta + 60 ) f_6(\phi) = 0 \; , \ee
ruling out the possibility of $E_{7(7)}$ invariance for the unique 1/8 BPS $SU(8)$-invariant $\partial^{6} R^4$ type counterterm as well. 

From the uniqueness of these three BPS $SU(8)$-invariant $D=4$ operators, we also get constraints on the existence of duality-invariant counterterms in dimensions $D>4$. The unique forms of the functions $f_n(\phi)$ in Eq.\ \eqref{dualitycond} for each of the cases $n=3,\ 5,\ 6$ imply in turn

\bea  \frac{D-4}{D-2} 3 ( 29- D   ) = 42 \quad  &\Rightarrow& \quad  ( D - 8 ) ( D - 11) = 0  \\
\frac{D-4}{D-2} 5 ( 27- D   ) = 60 \quad  &\Rightarrow& \quad  ( D - 7 ) ( D - 12) = 0  \\
\frac{D-4}{D-2} 6 ( 26- D   ) = 60 \quad  &\Rightarrow& \quad  ( D - 6 ) ( D - 14) = 0 \ . 
\eea

 From the $n=3$ case, we learn that duality-invariant $R^4$ type operators are possible in $D=11$ and $D=8$, as we have already seen in Eq.\ \eq{D8invariant}, but we also learn that duality-invariant $R^4$ type operators cannot exist in other dimensions.

 From the $n=5$ and $n=6$ cases, we learn that duality-invariant $\partial^4 R^4$ and $\partial^6 R^4$ type counterterms can exist only in the already-known dimensions $D=7$ and $D=6$, respectively. In particular, this rules out the existence of an $E_{6(6)}$ invariant counterterm for a putative 4-loop divergence in five dimensions. This explains the absence of the $D=5$, $L=4$ divergence computed explicitly in \cite{Bern:2009kd}. This also shows that there are no invariants of types $\partial^4 R^4$ or $\partial^6 R^4$ in eleven dimensions, although the higher-order supersymmetrisation of the $R^4$ invariant could involve a $\partial^6 R^4$ term. 

For completeness, let us discuss briefly the case of type II supergravity in ten dimensions. We define $\phi_{\scriptscriptstyle 10}$ as in (\ref{KKmetric}), and let $\phi$ be the ten-dimensional dilaton (this discussion is valid for both type IIA and type IIB). The existence of an invariant at order $\alpha^{\prime \, n}$ and $\ell$ loops in string theory with respect to the classical supersymmetry transformations would imply the following Laplace equation for the corresponding $n$-loop four dimensional $f_n(\phi)$: 
\bea\Delta   \, e^{- 6n \phi_{\scriptscriptstyle 10}Ê+ \left( 2\ell - \frac{n}{2} \right) \phi  } &=& \left( \frac{1}{48} \frac{Ê\partial^2 \, }{\partial \phi_{\scriptscriptstyle 10}{}^2 } + \frac{11}{4}Ê \frac{\partial\, }{\partial \phi_{\scriptscriptstyle 10} } +  \frac{Ê\partial^2 \, }{\partial \phi^2 } + Ê \frac{\partial\, }{\partial \phi } \right) e^{- 6n \phi_{\scriptscriptstyle 10}Ê+ \left( 2\ell - \frac{n}{2} \right) \phi  }\nn \\ &=& \left( n (n-17) + 4\ell \Bigl( \ell - \frac{n-1}{2} \Bigr) \right) \, e^{- 6n \phi_{\scriptscriptstyle 10}Ê+ \left( 2\ell - \frac{n}{2} \right) \phi  } \; . \eea
Consistently with the contributions to the string-theory effective action; the only possible invariants at order $\alpha^{\prime \, 3}$ appear at the tree level and one loop, the only possible invariants at order $\alpha^{\prime \, 5}$ appear at the tree level and two loops; and the only possible invariant at order $\alpha^{\prime \, 6}$ appears at three loops. This is not in contradiction with the 0, 1 and 2-loop contributions to the effective action at order $\alpha^{\prime \, 6}$, because they define the order $\alpha^{\prime \, 6}$ parts of the $\ell=0$ and $\ell=1$ $\alpha^{\prime \, 3}$ invariants.

Although our purely field-theoretic discussion is completely different in nature from the one given in terms of the non-perturbative string theory effective action in \cite{Green:2010sp}, there are some similarities that are worth pointing out. The Laplace equation \eq{dualitycond} that we have shown to be required for the scalar prefactor functions multiplying the $\partial^{2(n-3)} R^4$ terms in $D=4$ BPS candidate counterterms has been derived in \cite{Green:2005ba} by an analysis of the expected properties of U-duality-invariant functions $\cE_n(\phi)$ multiplying the  $\partial^{2(n-3)} R^4$ term in the non-perturbative string theory effective action. This is rather natural, because it was conjectured in \cite{Green:2005ba} that this equation is implied by supersymmetry. Note, nevertheless, that the equations obtained in \cite{Green:2010sp} differ from ours in that we do not have Dirac delta-function source terms, which only appear when considering non-analytic components of the non-perturbative functions $\cE_n(\phi)$. We also do not consider the source for the 6-loop equation quadratic in the 3-loop function, since we are just interested in supersymmetry invariants with respect to the tree-level supersymmetry transformations when searching for candidate counterterms for a first logarithmic divergence. Relying on the uniqueness of the BPS invariants in four dimensions, we have been able to demonstrate that the possible supergravity logarithmic divergences can occur only when the Laplace equation satisfied by the threshold function multiplying the $\partial^{2(n-3)} R^4$ term is $ \Delta_{\scriptscriptstyle D} \, f_n(\phi) = 0 $, in which case $ f_n(\phi) = 1$ is a solution and a duality-invariant counterterm can exist. This occurs precisely when the Laplace equation satisfied by the non-perturbative function $\cE_n(\phi)$ admits a Dirac delta-function source \cite{Green:2010sp}. It is precisely these Dirac delta-function sources which imply that $\cE_n(\phi)$ involves a logarithm of the effective dilaton, which would be associated to a logarithm of the Mandelstam invariant in the supergravity limit, and thus to a logarithmic divergence in field theory. 

 To summarise, the results of this analysis for the maximal $D=4$ supergravity theory are that the 1/2, 1/4 and 1/8 BPS candidate counterterms are ruled out by the requirement of continuous $E_{7(7)}$ duality invariance. Consequently, the first viable $D=4$ candidate counterterm with $E_{7(7)}$ invariance will be the non-BPS $\partial^8R^4$ operator anticipated at seven loops \cite{Howe:1980th}. In superspace language, this first $E_{7(7)}$ invariant candidate is simply the volume of superspace, $\int d^4x d^{32}\theta\, \textrm{det} E$. It remains to be verified whether this invariant is non-vanishing subject to the classical field equations of the $N=8$ theory.

In addition, we have shown that the absence of an $E_{6(6)}$ invariant $\partial^6 R^4$ counterterm in $D=5$ explains the 4-loop finiteness of the maximal supergravity theory in five dimensions. 


\section{Harmonic measures}


Another aspect of the difficulty in constructing non-linear invariants in maximal supergravity is that the necessary measures that generalise the linearised ones do not always exist. Here we discuss this issue in the case of $D=4, N=8$ supergravity. At the linearised level, there are three short invariants which can be written as integrals over certain harmonic superspaces \cite{Drummond:2003ex}. We briefly review these and then discuss how one might try to generalise these integrals to the non-linear case. 

We recall that harmonic superspace is the product of ordinary superspace with a coset of the R-symmetry group $G$ which is always chosen to be a compact complex manifold, $K$ \cite{Rosly:1982,Galperin:1984av,Karlhede:1984vr}. Instead of working on $K$ directly, it is convenient to work with fields that are defined on $G$ and then demanding that their dependence on the isotropy group $H$ defining $K$, $K=H\bsh G$, be fixed in such a way that they are equivalent to tensor fields on $K$ \cite{Galperin:1984av}. We shall denote an element of $G$ by $u_I{}^i$ where $G$ ($H$) acts to the right (left) on the small (capital) index, and its inverse by $v_i{}^I$. In flat $D=4$ superspace the derivatives are $(\del_a, D_{\a i}\bar D_{\adt}^i),\ i=1,\ldots N$. The introduction of the new variables allows us to define subsets of the odd derivatives that mutually anticommute without breaking the R-symmetry. Such a subset with $p$ $D$s and $q$ $\bar D$s is called a Grassmann (G)-analytic structure of type $(p,q)$, and a G-analytic field of type $(p,q)$ is one that is annihilated by all of these derivatives. 

For $N=8$ we can take $H=S(U(p)\xz U(8-(p+q))\xz U(q)$ and set $u_I{}^i=(u_r{}^i, u_{R}{}^i,u_{r'}{}^i)$. The $(p,q)$ mutually anticommuting derivatives are

\be
D_{\a r}:=u_r{}^i D_{\a i}  \qquad {\rm and}\quad \bar D_{\adt}^{r'}:=\bar D_{\adt}^i v_i{}^{r'}\ ,
\label{1}
\ee
for $r=1,\cdots p$ and $r'=(N-q) , \cdots N$. 
As the superfields will also depend on $u$ we need to introduce derivatives on $SU(8)$; they are the right-invariant vector fields $D_I{}^J$ and they satisfy the Lie algebra relations  of $\gs\gu(8)$. Their action on the $u,v$ variables is given by

\be
D_I{}^J u_K{}^k=\d_K{}^J u_I{}^k-\frac{1}{8} \d_I{}^J u_K{}^k;\qquad D_I{}^J v_k{}^K=- \d_I{}^K v_k{}^J+\frac{1}{8} \d_I{}^J v_k{}^K\ .
\label{2}
\ee

The derivatives split into subsets: $(D_r{}^s, D_R{}^S, D_{r'}{}^{s'})$ correspond to the isotropy subalgebra while $(D_r{}^S, D_r{}^{s'}, D_R{}^{s'})$ can be thought of as the components of the $\bar\del$-operator on the complex manifold $K$. The remaining derivatives are the complex conjugates of these. This means that we can have superfields that are G-analytic (annihilated by $(D_{\a r},\bar D_{\adt}^{r'})$), superfields that are harmonic, or H-analytic (annihilated by $(D_r{}^S, D_r{}^{s'}, D_R{}^{s'})$, and superfields that are annihilated by both sets since they are compatible in the sense that they are closed under graded commutation. We shall call such superfields analytic. They are the integrands for the short invariants. The fact that they are H-analytic implies that they have short expansions in $u$, because $K$ is compact as well as complex, and the fact that they are G-analytic means that they can be integrated over $32-2(p+q)$ odd coordinates rather than the full 32. 

The $N=8$ field strength superfield $W_{ijkl}$ is in the 70 of $SU(8)$; it is totally antisymmetric and self-dual on its $SU(8)$ indices and satisfies

\bea
D_{\a i} W_{jklm}&=&D_{\a [i} W_{jklm]}\nn\w1
\bar D_{\adt}^i W_{jklm}&=&-\frac{4}{5}\d^i_{[j} \bar D_{\adt}^n W_{klm]n}\ .
\label{3}
\eea

The $R^4$ invariant can be written in $(4,4)$ superspace. The field $W:=\frac{1}{4!} \ve^{rstu} u_r{}^i\ldots u_u{}^l W_{ijkl}$ is easily seen to be G-analytic and is also obviously H-analytic on the coset $S(U(4)\xz U(4))\bsh SU(8)$. It is preferable to write $W$ as $W_{1234}$ as this exhibits the charges explicitly. The $R^4$ invariant is

\be
I=\int\, d^4 x\, du\, [D_5\ldots D_8 \bar D^1\ldots\bar D^4]^2 \, (W_{1234})^4
\label{4}
\ee

where $du$ denotes the standard measure on the coset and the theta-integration is represented as differentiation with respect to all of the spinorial derivatives that do not annihilate $W$.  It is easily seen to be unique as it makes use of  the only dimension-zero analytic integrand with the right charges.
The other two short invariants $\partial^4 R^4$ and $\partial^6 R^4$ can be written in a similarly unique  fashion as integrals over $(2,2)$ and $(1,1)$ harmonic superspaces respectively. 

We now want to try to generalise this picture to curved superspace.\footnote{$N=2$ curved harmonic superspace was first studied in \cite{Galperin:1987ek}; the sort of analysis given here was described for $N\leq 4$ conformal supergravity theories in \cite{Hartwell:1994rp}.} In superspace the tangent spaces split invariantly into even and odd sectors (there is no supersymmetry in the tangent space) and for $N=8$ the structure group is $SL(2,\bbC)\xz SU(8)$. Because of the split structure, it is always best to work in a preferred basis. The preferred basis one-forms are related to the coordinate one-forms by the supervielbein, $E^A=dz^M E_M{}^A$; their duals are denoted $E_A$. We set $E^A=(E^a, E^{\a i},\bar E^{\adt}_i)$, where $a$ is a vector index. $SL(2,\bbC)$ acts on the spinor indices $\a,\adt$ and also on the vector index $a$ via the corresponding element of the Lorentz group, while the local $SU(8)$ acts on $i,j$, etc. We also have a set of connection one-forms $\O_A{}^B$ with

\bea
\O_{\a i}{}^{\b j}&=&\d_\a{}^\b \O_i{}^j + \d_i{}^j \O_\a{}^\b\nn\w1
\O_{ab}\rightarrow \O_{\a\adt,\b\bdt}&=&\ve_{\adt\bdt} \O_{\a\b} + \ve_{\a\b} \bar\O_{\adt\bdt}\ ,
\label{5}
\eea

where  we have used the usual relation between vector indices and pairs of spinor indices. $\O_{\adt j}^{i \bdt}$ is the complex conjugate of $\O_{\a i}{}^{\b j}$ and the off-diagonal elements of $\O_A{}^B$ are zero. The torsion and curvature tensors are defined in the usual way, using the covariant exterior derivative $D$, by

\be
T^A=D E^A;\qquad R_A{}^B=d\O_A{}^B + \O_A{}^C \O_C{}^B\ .
\label{6}
\ee

In $N=8$ supergravity, the scalars are described by an element $\cV$ of the group $E_{7(7)}$ \cite{Cremmer:1979up}, where the local $SU(8)$ acts from the left and the rigid $E_{7(7)}$ acts from the right. The Maurer-Cartan form $\F$ is

\be
\F=d \cV \cV^{-1}=P+Q \ ,
\label{7}
\ee

where $P$ is in the $70$ of $SU(8)$ and $Q$ is the $\gs\gu(8)$ connection which is to be identified with $\O_i{}^j$ above. In the geometrical quantities, the scalars only appear through the vector part of the one-form $P$, \ie $P_a$, which one can think of as a suitably defined pullback of the covariant derivative for the scalar target manifold.

The constraints on the various tensors that need to be imposed in order to describe on-shell $N=8$ supergravity can be found in \cite{Brink:1979nt,Howe:1981gz}. At dimension zero, the only non-vanishing torsion is

\be
T_{\a i,\bdt}^{\  \  \ j  \,c}=-i\d_i{}^j (\s^c)_{\a\bdt}\ ,
\label{8}
\ee

and the only non-vanishing dimension one-half torsion is

\be
T_{\a i,\b j, k}^{\ \ \ \ \ \cdt}=\ve_{\a\b} \bar \chi^{\cdt}_{ijk}
\label{9}
\ee

and its conjugate, where $\chi_\a^{ijk}$ is the superfield whose leading component is the physical spinor field that transforms under the $56$ of $SU(8)$.

This brief outline is enough to enable us to discuss whether there can be harmonic superspace measures of the required type in the non-linear theory where the $SU(8)$ R-symmetry becomes local.  We need to enlarge the superspace by adjoining some group variables $u$. The resulting space is the principal bundle associated with the $SU(8)$ part of the structure group and harmonic superspace is the associated bundle with typical fibre $K=H\bsh SU(8)$, for the appropriate isotropy group $H$. The idea is to search for appropriate CR structures, that is, complex, involutive  distributions which involve $2(p+q)$ odd directions and the accompanying holomorphic structures in the bundle coordinates. The way to do this is to introduce the horizontal lift basis in the total space of the bundle corresponding to a preferred basis in the base manifold. We have

\be
\tE_A:=E_A -\O_{A I}{}^J D_J{}^I\ ,
\label{10}
\ee

where one switches to an $I$ index from an $i$ index by means of $u_I{}^i$ and its inverse, as in the flat case. We then have

\be
[\tE_{\a I},\tE_{\b J}]= -T_{\a I\b J}{}^C \tE_C + R_{\a I\b J K}{}^L D_L{}^K\  + \ldots,
\label{11}
\ee

where the additional terms are irrelevant for this discussion, and similarly for the dotted and mixed commutators of the spinorial lifted bases. Now suppose that we require the CR structure to include  $\tE_{\a r},\  r=1\ldots p$ and $\tE_{\bdt}^{s'},\ s'=N-q,\ldots N$. We can see immediately that this leads to consistency conditions on the dimension one-half torsion, namely

\be
T_{\a r,\b s, \cdt t}=T_{\a r,\b s, \cdt T}=0\ ,
\label{12}
\ee

since otherwise these derivatives would not close among themselves. From the explicit form of the dimension one-half torsion (\ref{9}) we can see that the only possibility is that $r$ can only take on one value. A similar result holds for the dotted indices, and so we conclude that we can only have Grassmann analyticity of type $(1,1)$ in the full theory. There are also conditions on the curvature; for example, one must have $R_{\a 1,\b 1,1}{}^8=0$. From the explicit expressions for the dimension-one curvature given in  \cite{Brink:1979nt,Howe:1981gz}, one can see that these conditions are indeed satisfied. One can also check that it is possible to have $(1,1)$ analytic fields carrying $U(1)$ charges in pairs of the type  $_1^8$; this is necessary if there are to be integrands with the right charges that can be integrated with respect to the $(1,1)$ measure. As a CR structure is necessary in order that we can have harmonic superspaces with fewer odd coordinates (also called analytic superspaces) it follows that harmonic measures do not exist for $(p,q)=(4,4)$ and $(2,2)$ G-analyticity, and therefore that there can be no straightforward generalisation of the $R^4$ and $\partial^4 R^4$ invariants, expressed as harmonic superspace integrals, to the full non-linear theory that are compatible with local $SU(8)$ symmetry.

In the case of $(1,1)$ analyticity, relevant to the $\partial^6 R^4$ invariant, the measure should exist, which suggests that this invariant can be written as an harmonic superspace integral. However, the harmonic measure is definitely not R-symmetric, which implies that the integrand must be a non-trivial function of the scalars $\cV$. In the formulation with gauged $SU(8)$ and linearly realised rigid $E_{7(7)}$, the measure will be $E_{7(7)}$ invariant whereas the integrand will necessarily transform non-trivially with respect to $E_{7(7)}$. It would then follow that the $\partial^6 R^4$ invariant is not $E_{7(7)}$ invariant, in agreement with the conclusion of the preceding section. 

Note that this is not in contradiction with the existence of BPS duality invariants in higher dimensions (such as $R^4$ in $D=8$, $\partial^4 R^4$ in $D=7$ and $\partial^6 R^4$ in $D=6$), since the BPS invariants are not unique in dimensions $D > 5$.

The non-existence of harmonic measures for the 1/2 and the 1/4 BPS invariants is not in contradiction with the existence of these non-linear invariants in the full non-linear theory. Indeed as we will discuss in the next section, not all supersymmetry invariants can be written as harmonic superspace integrals, and some are only described in terms of closed super-D-form.


\section{Invariants as closed super-four-forms}


 An alternative approach to the construction of superinvariants is afforded by the ectoplasm formalism \cite{Voronov,Gates:1997kr,Gates:1997ag}. In $D$-dimensional spacetime, consider a closed super-$D$-form, $L_D$, in the corresponding superspace. The integral of the purely bosonic part of this form over spacetime is guaranteed to be supersymmetric by virtue of the closure property. Moreover, if $L_D$ is exact it will clearly give a total derivative so that we are really interested in the $D$th superspace cohomology group. As we have seen in the preceding section, one cannot define a harmonic measure for every invariant, and in particular, not for the 1/2 and 1/4 BPS invariants in $N=8$ supergravity. However, according to the algebraic Poincar\'e Lemma, any supersymmetry invariant necessarily defines a closed super-$D$-form.

The analysis below is a little involved, but the basic ideas are easy to state. The non-vanishing components of a closed super-four-form in $N=8$ superspace can define two types of cocycle: those that correspond directly to (sub)superspace integrals and those that do not. The former will correspond to allowed measures, and so, of the short invariants, only the one-eighth BPS one can be expected to be of this type. The other two will not, and hence will have different types of cocycle to those of full superspace integrals; we can thus expect these invariants to be protected by virtue of the algebraic renormalisation procedure.  The explicit expressions for the closed BPS super-four-forms are difficult to construct and we will not attempt to do this here. We can, however, obtain information about $E_7$ invariance by studying the linearised approximation. We shall show that the linearised cocycles for $R^4$ in $N=8$ and $N=5,6$ and $\del^2 R^4$ in $N=6$ are not invariant under linearised duality transformations (constant shifts of the scalars) so that the full invariants cannot possibly be.   Indeed, the linearised four-point contribution to the lowest component of the super-four-form contributes directly to the spacetime invariant at eight points as can be seen from \eq{Lagrange} below.

In order to analyse superspace cohomology, it is convenient to split forms into their even and odd parts. Thus  a $(p,q)$-form is a form with $p$ even and $q$ odd indices, totally antisymmetric on the former and totally symmetric on the latter. The exterior derivative can likewise be decomposed into parts with different bi-degrees,

\be
d=d_0+ d_1 + t_0 +t_1\ ,
\label{13}
\ee

where the bi-degrees are $(1,0)$, $(0,1)$, $(-1,2)$ and $(2,-1)$ respectively. So $d_0$ and $d_1$ are basically even and odd derivatives, while $t_0$ and $t_1$ are algebraic. The former acts by contracting an even index with the vector index on the dimension-zero torsion and then by symmetrising over all of the odd indices. The equation $d^2=0$ also splits into various parts of which the most relevant components are

\be
t_0^2=0; \qquad d_1 t_0 + t_0 d_1 =0; \qquad  d_1^2 + t_0 d_0 + d_0 t_0=0\ .
\label{14}
\ee

The first of these equations allows us to define $t_0$-cohomology groups, $H_t^{p,q}$ \cite{Bonora:1986ix}, and the other two allow us to introduce the spinorial derivative $d_s$ which maps $H_t^{p,q}$ to $H_t^{p,q+1}$ by $d_s [\o_{p,q}]=[d_1 \o_{p,q}]$, where the brackets denote $H_t$ cohomology classes. This also squares to zero, and hence allows one to define spinorial cohomology groups $H_s^{p,q}$ \cite{Cederwall:2001dx,Howe:2003cy}. The point of this is that one can often generate closed super-$D$-forms from elements of these cohomology groups.

In the context of curved superspace it is important to note that the invariant is constructed from the top component in a coordinate basis,

\be
I= \frac{1}{D!}  \int\, d^D x\;  \varepsilon^{m_D\dots m_1} \; E_{m_D}{}^{A_D} \cdots E_{m_1}{}^{A_1}  \;  L_{A_1 \dots A_D} (x,\th=0)\ .
\label{15}
\ee

One transforms to a preferred basis by means of the supervielbein $E_M{}^A$. At $\th=0$ we can identify $E_m{}^a$ with the spacetime vielbein $e_m{}^a$ and $E_m{}^{\ua}$ with the gravitino field $\psi_m{}^{\ua}$ (where $\ua$ includes both space-time $\a,\, \dot{\a}$ and internal $i$ indices for $N=8$). In four dimensions, we therefore have 

\begin{multline} 
I=\frac{1}{24} \int\,\bigl(  e^a_{\,\wedge} e^b_{\, \wedge} e^c_{\, \wedge} e^d \, \, L_{abcd} + 4 e^a_{\,\wedge} e^b_{\, \wedge} e^c_{\, \wedge} \psi^{\ua}  L_{abc\ua} + 6 e^a_{\,\wedge} e^b_{\, \wedge} \psi^\ua_{\, \wedge} \psi^\ub \, \, L_{ab\ua\ub} \bigr . \\ \bigl .  + 4 e^a_{\,\wedge} \psi^\ua_{\, \wedge} \psi^\ub_{\, \wedge} \psi^{\uc}  L_{a\ua\ub\uc} +  \psi^\ua_{\, \wedge} \psi^\ub_{\, \wedge} \psi^{\uc}_{\,\wedge} \psi^\ud  L_{\ua\ub\uc\ud} \bigr)  \; . \label{Lagrange}
\end{multline}

By definition, each component $L_{abcd},\, L_{abc\ua} ,\, L_{ab\ua\ub},\, L_{a\ua\ub\uc}, \, L_{\ua\ub\uc\ud}$ is supercovariant at $\theta=0$. This is a useful formula because one can directly read off the invariant in components in this basis. 

In $N=8$ supergravity, all the non-trivial $t_0$-cohomology classes lie in $H_t^{0,4}$. Invariants are therefore completely determined by their $(0,4)$ components $L_{\ua\ub\uc\ud}$, and all non-trivial $L_{0,4}$ satisfying $[d_1 L_{0,4}]=0$ in $t_0$-cohomology (\ie that give rise to non-trivial elements of $H_s^{0,4}$) define non-trivial invariants. $H_t^{0,4}$ is the set of functions of fields in the symmetric tensor product of four ${\bf 2} \times {\bf 8} \oplus \overline{\bf 2} \times \overline{\bf 8}$ of $SL(2,\bbC) \times SU(8)$ without $SU(8)$ contractions (since such functions would then be $t_0$-exact). Because of the reducibility of the representation, it will be convenient to decompose  $L_{\ua\ub\uc\ud}$ into components of degree $(0,p,q)$ ($p+q=4$) with $p$ ${\bf 2} \times {\bf 8}$ and $q$ $\overline{\bf 2} \times \overline{\bf 8}$ symmetrised indices. 

We will classify the elements of $H_t^{0,4}$ into three generations.\footnote{There is also an element of $H_t^{0,4}$ of degree $(0,2,2)$ in the $[0,0|0100010]$ representation. This seems unlikely to  play any role and we shall not discuss it further here.} The first generation corresponds to elements that lie in the antisymmetric product of four ${\bf 2} \times {\bf 8} \oplus \overline{\bf 2} \times \overline{\bf 8}$ of $SL(2,\bbC) \times SU(8)$, and can therefore be directly related to the top component $L_{4,0}$ through the action of the superderivatives. We will write $M_{0,p,q}$ for the corresponding components of a given $L_{0,4}$. They lie in the following irreducible representations of $SL(2,\bbC) \times SU(8)$:

\be\begin{split}  
M_{0,4,0} &: [0,0|0200000] \\
M_{0,3,1} &: [1,1|1100001] \\
M_{0,2,2} &: [2,0|2000010] 
\end{split}\hspace{15mm}\begin{split}
\bar M_{0,0,4} &: [0,0|0000020] \\
\bar M_{0,1,3} &: [1,1|1000011] \\
\bar M_{0,2,2} &: [0,2|0100002] \; . 
\end{split}\ee

In order to understand the constraints that these functions must satisfy in order for $L_{0,4}$ to satisfy the descent equation 

\be [d_1 L_{0,4}]=0 \; ,  \label{Descent} \ee

it is useful to look at the possible representations of $d_1 L_{0,4}$ which define $H_t^{0,5}$ cohomology classes in general, without assuming any {\it \`a priori} constraint. We will split $d_1 = d_{1,0} + d_{0,1}$ according to the irreducible representations of $SL(2,\bbC) \times SU(8)$. One computes that 

\be\begin{split}  
[d_{1,0} M_{0,4,0}] &: [1,0|1200000] \\
[d_{0,1} M_{0,4,0}] &: [0,1|0200001] \\
[d_{1,0} M_{0,3,1}] &: [0,1|0200001] \oplus [2,0|2100001] \\
[d_{0,1} M_{0,3,1}] &: [1,0|1100010] \oplus [1,2|1100002] \\
[d_{1,0} M_{0,2,2}] &: [1,0|1100010] \oplus [3,0|3000010] \\
[d_{0,1} M_{0,2,2}] &: [2,1|2000011] 
\end{split}\hspace{15mm}\begin{split}
[d_{0,1} \bar M_{0,0,4}] &: [0,1|0000021] \\
[d_{1,0} \bar M_{0,0,4}] &: [1,0|1000020] \\
[d_{0,1} \bar M_{0,1,3}] &: [1,0|1000020] \oplus [0,2|1000012] \\
[d_{1,0} \bar M_{0,3,1}] &: [0,1|0100011] \oplus [2,1|2000011] \\
[d_{0,1} \bar M_{0,2,2}] &: [0,1|0100011] \oplus [0,3|0100003] \\
[d_{1,0} \bar M_{0,2,2}] &: [1,2|1100002] \; . 
\end{split}\ee

In order for the component 

\be L_{0,4}  = \sum_{p=0}^2  \bigl( M_{0,4-p,p} + \bar M_{0,p,4-p}  \bigr) \ee

to satisfy the descent equation (\ref{Descent}), the components $d_1 M_{0,p,q}$ must individually vanish in the $ [1,0|1200000],\, [2,0|2100001],\,  [3,0|3000010] $ representations and their complex conjugates, and their components in the $[0,1|0200001],\, [1,0|1100010],\,  [1,2|1100002]$ and their complex conjugates must cancel each other. This will indeed be the case if the invariant in question can be defined as a superaction and all the components of $L_{0,4}$ descend from a primary operator satisfying the appropriate constraint. However, as we have seen in the preceding section, there is no harmonic measure for the 1/2 and 1/4 BPS invariants, and this situation is therefore not the most general. 
 
What will happen for these invariants is that, although the components of $d_1  M_{0,p,q}$ in the $[0,1|0200001],\, [1,0|1100010] $ and their complex conjugate representations cancel each other, the components in the  $ [1,0|1200000],\, [2,0|2100001],\,  [3,0|3000010],\,  [1,2|1100002] $ and the corresponding complex conjugates will not vanish. The latter will nevertheless be cancelled by the $d_1$ variation of a second generation of functions $N_{0,p,q}$ in  $H_t^{0,4}$,

\begin{gather}\begin{split}  
N_{0,4,0} &: [2,0|2100000] \\
N_{0,3,1} &: [3,1|3000001] 
\end{split}\hspace{15mm}\begin{split}
\bar N_{0,0,4} &: [0,2|0000012] \\
\bar N_{0,1,3} &: [1,3|1000003] 
\end{split} \nn \\
N_{0,2,2} : [2,2|2000002] \;  . \end{gather}

Indeed, one computes that the components of $[d_1 N_{0,p,q}]$ lie in the following representations 

\be\begin{split}  
[d_{1,0} N_{0,4,0}] &: [1,0|1200000] \oplus [3,0|3100000]\\
[d_{0,1} N_{0,4,0}] &: [2,0|2100001] \\
[d_{1,0} N_{0,3,1}] &: [2,0|2100001] \oplus [4,1|4000001] \\
[d_{0,1} N_{0,3,1}] &: [3,0|3000010] \oplus [3,2|3000002] \\
[d_{1,0} N_{0,2,2}] &: [1,2|1100002] \oplus [3,2|3000002]
\end{split}\hspace{15mm}\begin{split}
[d_{0,1} \bar N_{0,0,4}] &: [0,1|0000021] \oplus [0,3|0000013] \\
[d_{1,0} \bar N_{0,0,4}] &: [0,2|1000012] \\
[d_{0,1} \bar N_{0,1,3}] &: [0,2|1000012] \oplus [1,4|1000004] \\
[d_{1,0} \bar N_{0,3,1}] &: [0,3|0100003] \oplus [2,3|2000003] \\
[d_{0,1}  N_{0,2,2}] &:  [2,1|2000011] \oplus [2,3|2000003] \; . 
\end{split}\ee

In addition to cancelling the components $[ d_1  M_{0,p,q} ]$, the components $[d_1 N_{p,q}]$ must cancel each other in the $ [3,2|3000002]$ representation and its complex conjugate. Then there are two possibilities: either the components of $[d_1 N_{p,q}]$ identically vanish in the $[3,0|3100000]$, the $[4,1|4000001]$ and their complex conjugates, or a third generation of $O_{0,4,0}$ functions and their $\bar O_{0,0,4}$ complex conjugates in $H_t^{0,4}$ is required to cancel them,

\be O_{0,4,0} : [4,0|4000000] \hspace{10mm} \bar O_{0,0,4} : [0,4|0000004] \; . \ee

Now, $[d_1 O_{0,4,0}]$ lies in the following representations of $H_t^{0,5}$

\be\begin{split}  
[d_{1,0} O_{0,4,0}] &: [3,0|3100000] \oplus [5,0|5000000]\\
[d_{0,1} O_{0,4,0}] &: [4,1|4000001] 
\end{split}\hspace{15mm}\begin{split}
[d_{0,1} \bar O_{0,0,4}] &: [0,3|0000013] \oplus [0,5|0000005] \\
[d_{1,0} \bar O_{0,0,4}] &: [1,4|1000004] \; , 
\end{split}\ee

and in addition to cancelling $[d_1 N_{p,q}]$ in the $[3,0|3100000]$, the $[4,1|4000001]$ and their complex conjugates, the components of $d_{1,0} O_{0,4,0}$ in the $[5,0|5000000]$ must identically vanish. 

To conclude this discussion, we have seen from the $t_0$-cohomology analysis that there exist more general cocycle structures than those associated to invariants that can be written as (harmonic) superspace integrals. The absence of harmonic measures for the 1/2 and 1/4 BPS invariants is therefore not in contradiction with the existence of such invariants. However, their cocycle structures involve two or three supermultiplets instead of only one, corresponding to the second generation of operators $N_{0,p,q}$, and possibly the third $O_{0,4,0}$. The expectation is that the 1/2 BPS invariant will admit a cocycle involving three generations, 

\be L_{0,4}^\demi =  \sum_{p=0}^2  \bigl( M^\demi_{0,4-p,p} + \bar M^\demi_{0,p,4-p}  \bigr) + \sum_{p=0}^1  \bigl( N^\demi_{0,4-p,p} + \bar N^\demi_{0,p,4-p}  \bigr) + N^\demi_{0,2,2}  + O^\demi_{0,4,0} + \bar O^\demi_{0,0,4} \; ,  \ee

and the 1/4 BPS invariant will admit a cocycle involving two generations, 

\be  L_{0,4}^\quart =  \sum_{p=0}^2  \bigl( M^\quart_{0,4-p,p} + \bar M^\quart_{0,p,4-p}  \bigr) + \sum_{p=0}^1  \bigl( N^\quart_{0,4-p,p} + \bar N^\quart_{0,p,4-p}  \bigr) + N^\quart_{0,2,2}    \; . \ee

We have not derived the explicit functions which define these cocycles, but we would like to point out that the $F^4$ invariants in super Yang--Mills theory in ten dimensions define explicit example of such cocycles involving several generations of $t_0$-cohomology classes \cite{bhlsw}. From this perspective, it seems that a careful study of the implications of supersymmetry Ward identities within the algebraic approach should rule out the possibility of both the 3 and 5-loop logarithmic divergences in $N=8$ supergravity.  (We recall also that the 4-loop divergence has no available on-shell nonvanishing counterterm \cite{Drummond:2003ex}.) However, the existence of a 1/8 BPS harmonic measure suggests that the 1/8 BPS cocycle has the same structure as the cocycle associated to full superspace integral invariants, and therefore that the supersymmetry Ward identities alone will be unable to rule out the corresponding 6-loop divergence within the algebraic approach. However, as we have discussed in the preceding section, the integrand in that case must be a function of the scalar superfield, which implies that it cannot be $E_{7(7)}$ invariant, and therefore that the $E_{7(7)}$ Ward identities nonetheless rule out this divergence. 

The non-existence of a 1/2 BPS measure does not permit one to conclude directly that the $R^4$ invariant cannot be $E_{7(7)}$ invariant, without relying on the dimensional reduction argument presented in the first section. Nevertheless, it follows from the structure of the invariant (\ref{Lagrange}), that knowledge of the cocycle $L^\demi_4$ in the quartic field approximation provides information about terms of orders up to 8 in the invariant. If $I^\demi$ were invariant with respect to $E_{7(7)}$, then it would follow from the representation of $E_{7(7)}$ on the fields that each component $L^\demi_{abcd},\, L^\demi_{abc\ua}  ,\, L^\demi_{ab\ua\ub},\, L^\demi_{a\ua\ub\uc}, \, L^\demi_{\ua\ub\uc\ud}$ would independently have to be $E_{7(7)}$ invariant. In the linearised approximation, this means that each component would be invariant at lowest order with respect to a constant shift of the scalar superfield $W^{ijkl}$. It was pointed out in \cite{Howe:1981xy} that $L_{abcd}$ is shift invariant, but we shall see that the last component $L^\demi_{\ua\ub\uc\ud}$ is not, hence establishing that  $I^\demi$ is not fully $E_{7(7)}$ invariant. 

To start with, note that the 1/2 BPS invariant admits a superaction form in the linearised approximation. It follows that the second and third generations of $(0,4)$ components are not required in this approximation, and that $N^\demi_{0,p,q}$ and $O^\demi_{0,4,0}$ are at least quintic in fields. In order to establish the non-shift-invariance of $L^\demi_{0,4}$ in the quartic field approximation, it will be enough to consider its $M^\demi_{0,4,0}$ component. The latter can be obtained by acting on the 1/2 BPS primary operator defined by $W^4$ in the $[0004000]$ of $SU(8)$ with the $D^8$ in the $[0,0|0002000]$, and $\bar D^4$ in the $[0,0|0000020]$. With the conventional notation\footnote{That is, $F,\rho$ and $C$ are respectively the $(2,0), (3,0)$ and $(4,0)$ components of the spin one, three-halves and two field-strength tensors}

\be D_{\a p} W^{ijkl} = \delta_p^{[i} \chi^{jkl]}_\a \; , \quad D_{\a l} \chi^{ijk}_\b = \delta_l^{[i} F^{jk]}_{\a\b}
 \; , \quad D_{\a k} F^{ij}_{\b\c} = \delta_k^{[i} \rho^{j]}_{\a\b\c} \; , \quad D_{\a j} \rho^i_{\b\c\d} = \delta^i_j C_{\a\b\c\d} \; , \ee
 
one obtains that $D^8 W^4$ in the $[0,0|0002000]$ has the form 

\be D^8 W^4 \sim W^2 C^2 + W \chi \rho C + W F^2 C + W F \rho^2 + \chi^2 F C + \chi F^2 \rho + F^4 \; , \label{D8W4}  \ee

where the index contractions and symmetrisations are unambiguously determined by the representation. Since we are interested in the shift invariance of $M^\demi_{0,4,0}$, we can already disregard the three last terms. Applying finally $\bar D^4$ to (\ref{D8W4}), one obtains various terms linear in W, terms in $W^3 C ,\, W^2 F^2,\,   W^2 \chi \rho$ and $W \chi^2 F$ involving four derivatives, terms in $\bar \chi W \chi C$ and $\bar \chi W F \rho$ involving three derivatives, and terms in $\bar F W \rho^2$ and $W \bar F F C$ involving two derivatives. They are clearly all independent, taking into account the equations of motion, and one can discuss them separately. The term in $W^3 C$ is, for example, of the form 

\be W_{pqij} \partial^a \partial^b W^{pqrs} \partial^c \partial^d W_{klrs} C^{(+)}_{ac,bd}\ ,
\label{4.17} \ee

where $C^{(+)}$ denotes the self-dual part of the Weyl tensor, \ie $C_{\a\b\c\d}$ in spinor notation. The shift variation of \eq{4.17} is a total derivative, but it is clearly non-vanishing. Similarly, the terms in $W^2 F^2$ take the form 

\be \frac{1}{2} W^2 \partial^2  F \partial^2 F + W \partial W \partial F \partial^2 F  + W \partial \bigl(  \partial^2 W \partial F \partial F \bigr)  \; , \label{WWFF}  \ee

where the three terms involve one product of $W^2$ in the $[0002000]$ with $F^2$ in the $[0000020]$, one product of $W^2$ in the $[0010100]$ with $F^2$ in the $[0000101]$, while the third term moreover involves a product of $W^2$ in the $[0100010]$ with $F^2$ in the $[0001000]$. Once again, the shift variation of this set of terms is a non-vanishing total derivative. Hence, the shift variation of $M^\demi_{0,4,0}$ can be shown to be a non-vanishing total derivative. We recall that $E_{7(7)}$ invariance would be required separately for each of the $L_{(p,q)}$ forms in the complete invariant \eqref{Lagrange}, so at leading order each of these forms would need to be strictly shift invariant (total derivatives included) in order to achieve compatibility with $E_{7(7)}$.  Moreover, the structure of the 1/2 BPS supermultiplet implies that $M^\demi_{0,4,0}$ is uniquely determined from the primary operator $W^4$, and the 1/2 BPS cocycle does not admit other representatives, so no other terms could come to the rescue of the $E_{7(7)}$ symmetry.
 
We conclude that linearised analysis permits one to establish the $E_{7(7)}$ noninvariance of the full 1/2 BPS $R^4$ counterterm. However, this argument does not apply to the full 1/4 BPS $\partial^4 R^4$ counterterm. Indeed, one can define the 1/4 BPS counterterm in the linear approximation by acting with the 1/2 BPS measure on the non-primary 1/2 BPS quartic term $\partial_a W \partial_b W \partial^a W \partial^b W$ in the $[0004000]$ of $SU(8)$, which is manifestly shift invariant. But, of course, the shift invariance of the cocycle is a necessary but not sufficient condition for establishing  $E_{7(7)}$ invariance of the corresponding supersymmetry invariant, and the dimensional reduction argument of the first section shows indeed that it is not $E_{7(7)}$ invariant.


\section{$N=5,\, 6$ supergravity}

Note that the demonstration that $E_{7(7)}$ symmetry is preserved in perturbative theory for $N=8$ supergravity \cite{Bossard:2010dq}, generalises straightforwardly to the $N=5$ and $N=6$ cases for the duality symmetries $SU(5,1)$ and $SO^*(12)$ respectively, because all the one-loop $SL(2,\bbC)\times U(N)$ anomalies vanish \cite{Marcus}. Moreover, the linearised superalgebra in flat space can be embedded consistently into the corresponding superconformal algebra ${\mbox{\goth su}}(2,2 | N)$ similarly to the $N=8$ supergravity case, and one can again rely on superconformal representation analysis to prove that the BPS invariants are unique in these theories \cite{Drummond:2003ex}. In this section, we will show that analysis of the linearised super $4$-form associated to the corresponding $R^4$ invariants demonstrate that they also are not duality invariants, as in the $N=8$ supergravity case. We will correspondingly prove the absence of logarithmic divergences at three loops in these theories. Similarly, we will prove that the $\partial^2 R^4$ invariant is not $SO^*(12)$ invariant in $N=6$, incidentally proving that there is no logarithmic divergence at 4-loops.

In $N=6$ supergravity, the complex scalar superfield $W_{ij}$ and its complex conjugate $W^{ij}$ define the following multiplet by the recursive action of $D_{\a i}$: 
\bea D_{\a k} W_{ij} &=& \frac{1}{6} \varepsilon_{ijklmn} \chi^{lmn}_\a \; , \quad  D_{\a l} \chi^{ijk}_\b = \delta_l^{[i} F^{jk]}_{\a\b}
 \; , \quad D_{\a k} F^{ij}_{\b\c} = \delta_k^{[i} \rho^{j]}_{\a\b\c} \; , \quad D_{\a j} \rho^i_{\b\c\d} = \delta^i_j C_{\a\b\c\d} \; , \nn \\
D_{\a k} W^{ij} &=& \delta_k^{[i} \chi^{j]}_\a \; , \quad  D_{\a j} \chi^{i}_\b  = \delta_j^i F_{\a\b} \; . 
\eea

The linearised $R^4$ invariant can be obtained by acting with $\bar D^8 D^8$ in the $[0,0|02020]$ representation of $SL(2,\bbC) \times SU(6)$ on the 1/3 BPS operator $W_{ij} W_{kl} W^{pq} W^{mn}$ in the $[0,0|02020]$ representation.\footnote{We will not write explicitly the $U(1)$ weight, which is zero for both the measure and the integrand.} As for $N=8$ supergravity, the cocycle's last components are $M_{0,p,q}$ with 

\be\begin{split}  
M_{0,4,0} &: [0,0|02000] \\
M_{0,3,1} &: [1,1|11001] \\
M_{0,2,2} &: [2,0|20010] 
\end{split}\hspace{15mm}\begin{split}
\bar M_{0,0,4} &: [0,0|00020] \\
\bar M_{0,1,3} &: [1,1|10011] \\
\bar M_{0,2,2} &: [0,2|01002] \; , 
\end{split}\ee

and we will consider in particular the shift invariance of the $M_{0,4,0}$ component. The latter can be obtained by acting with $\bar D^4$ in the $[0,0|00020]$ and $D^8$ in the $[0,0|00020] $ on $W_{ij} W_{kl} W^{pq} W^{mn}$. $D^8 W^2 \bar W^2$ gives the $[0,0|00020]$ combination 

\be W^{ij} W^{kl} C^2 + W^{ij} \chi^{[k} \rho^{l]} C + W^{ij} F F^{kl} C + W^{ij} F \rho^{[k} \rho^{l]} + \dots \ee

where the dots stand for terms that are shift invariant. Applying then $\bar D^4$ to this expression, one obtains again various terms, including a single term in $W^3 C$ coming from $WF^2 C$ which reads

\be  \varepsilon_{ijpqrs} W^{pq} \partial^2 W^{rs}  \partial^2 W_{kl} C \; ,  \ee

projected into the $[0,0|02000]$ representation.  Similarly, one obtains various terms in $W^{ij} W^{kl}$  $F^{mn} F^{pq}$ which appear in combinations similar to (\ref{WWFF}) in $N=8$; as well as one term in $W^{ij} W_{kl} F F^{pq}$ coming from $W^{ij} F F^{kl} C$,

\be \varepsilon_{ijpqrs} W^{pq} F \partial^2 W_{kl} \partial^2 F^{rs} \ee

projected into the $[0,0|02000]$ representation. It follows that the result of a shift of the scalar field $W^{ij}$ in $M_{0,4,0}$ is non-vanishing, and not even a total derivative. We therefore conclude that the unique $R^4$ invariant in $N=6$ supergravity is not $SO^*(12)$ invariant. 

The $\partial^2 R^4$ counterterm can be obtained in a similar way from the 1/6 BPS operator $W^{ip} W^{jq} W_{kp} W_{lq}$ in the $[0,0|20002]$ representation, or from the non-primary 1/3 BPS operator $W_{ij}  W^{pq} \partial^a W_{kl} \partial_a W^{mn}$ in the $[0,0|02020]$. Note that any combination with two derivatives would necessarily be a total derivative in the $N=8$ theory because the scalar field is then real, which explains why there is no $\partial^2 R^4$ invariant in that case. All the possible ways of adding two derivatives to $W_{ij} W_{kl} W^{pq} W^{mn}$ are in fact equivalent, up to a total derivative. One can easily see that one cannot adjust the derivatives such that both $M_{0,4,0}$ and $\bar M_{0,0,4}$ are shift invariant. However, one must also consider the possibility of defining the cocycle directly from the 1/6 BPS operator $W^{ip} W^{jq} W_{kp} W_{lq}$. In that case $M_{0,4,0}$ is obtained by acting with $\bar D^6$ in the $[0,0|00200]$ and $D^{10}$ in the $[0,0|00002]$ on $W^{ip} W^{jq} W_{kp} W_{lq}$. Applying $D^{10}$, one already obtains an operator that does not depend on the scalars, so $M_{0,4,0}$ will be trivially shift invariant in this case. In order to exhibit the non-shift invariance of the 1/6 BPS cocycle, we must therefore consider other components. We will consider the $M_{0,2,2}$ component in the $[2,0|20010]$ of $SL(2,\bbC) \times SU(6)$. The latter can be obtained from the  1/6 BPS operator $W^{ip} W^{jq} W_{kp} W_{lq}$ by acting with $\bar D^{8}$ in the $[0,0|02000]$ and $D^{8}$ in the $[2,0|00101]$. The action of $\bar D^{8}$ gives a Lorentz scalar in the $[02000]$ 

\be W_{ij}ÊW_{kl} \bar C \bar C  + W_{ij} \bar \chi_{[k} \bar \rho_{l]} \bar C + W_{ij} \bar F \bar F_{kl} \bar C +  W_{ij} \bar F \bar\rho_{[k} \bar\rho_{l]} + \dots \ee

plus a Lorentz scalar in the $[21000]$

\be W_{i[j} \bar \chi_k \bar \rho_{l]} \bar C + \dots  \ee

where as before the dots stand for terms that are invariant with respect to a constant shift of the scalar fields, and the barred fields are the complex conjugate of the unbarred ones, which accordingly carry dotted Lorentz indices. The action of $ D^{8}$ on these terms is quite complicated, and we will focus on terms that are the most susceptible to fail shift invariance, namely the terms carrying two $W_{ij}$ not covered by derivatives. One can easily check that only the term $W^2 \bar C^2$ can produce such a term, and indeed produces one in the $[2,0|20010]$ 

\be W_{p(i} W_{j)q} \partial^4 \chi^p \partial^3 \bar \chi^{qkl}Ê\ee

where all the Lorentz indices of  $\partial^4 \chi^p$ in the $[5,4|00001]$ and $\partial^3 \bar \chi^{qkl}$ in the $[3,4|00100]$ are symmetrised, and are contracted such that $\partial^4 \chi^p \partial^3 \bar \chi^{qkl}$ is in the $[2,0|00101]$. The action of a constant shift of the scalar fields on this term gives a term in $W \partial^4 \chi \partial^3 \bar \chi$ which clearly cannot be compensated by other terms in $M_{0,2,2}$, although it might combine with other terms to give a non-vanishing total derivative. (But recall that, since this expression is still to be multiplied by gravitino factors, such a total derivative variation still constitutes duality non-invariance.)

Although the $\partial^2 R^4$ invariant admits several cocycle representatives in $N=6$ supergravity, none of them is invariant with respect to a constant shift of the scalar fields. We conclude that this candidate counterterm is not invariant with respect to the $SO^*(12)$ duality symmetry, and therefore that there is no logarithmic divergence at 4-loop in the theory. 

In $N=5$ supergravity, the complex scalar superfield $W_{i}$ and its complex conjugate $W^{i}$ define the following multiplet by the recursive action of $D_{\a i}$\,:

\bea D_{\a i} W_{j} &=& \chi_{\a ij} \; , \quad  D_{\a k} \chi_{\b ij} = \frac{1}{6} \varepsilon_{ijklp} F^{lp}_{\a\b}
 \; , \quad D_{\a k} F^{ij}_{\b\c} = \delta_k^{[i} \rho^{j]}_{\a\b\c} \; , \quad D_{\a j} \rho^i_{\b\c\d} = \delta^i_j C_{\a\b\c\d} \; , \nn \\
D_{\a j} W^{i} &=& \delta_j^{i} \chi_\a \; . 
 \eea
 
The linearised $R^4$ invariant can be obtained by acting with $\bar D^8 D^8$ in the $[0,0|2002]$ representation of $SL(2,\bbC) \times SU(5)$ on the 1/5 BPS operator $W_{i} W_{j} W^{k} W^{l}$ in the $[0,0|2002]$ representation. As for $N=8$ supergravity, the cocycle's last components are $M_{0,p,q}$ with 

\be\begin{split}  
M_{0,4,0} &: [0,0|0200] \\
M_{0,3,1} &: [1,1|1101] \\
M_{0,2,2} &: [2,0|2010] 
\end{split}\hspace{15mm}\begin{split}
\bar M_{0,0,4} &: [0,0|0020] \\
\bar M_{0,1,3} &: [1,1|1011] \\
\bar M_{0,2,2} &: [0,2|0102] \; , 
\end{split}\ee

and we will consider in particular the shift invariance of the $M_{0,4,0}$ component. The latter can be obtained by acting with $\bar D^4$ in the $[0,0|0020]$ and $D^8$ in the $[0,0|0002] $ on  $W_{i} W_{j} W^{k} W^{l}$. Evaluating $D^8 W^2 \bar W^2$ gives the $[0,0|0002]$ combination 

\be W^{i} W^{j} C^2 + W^{(i} \chi \rho^{j)} C  + \dots \ee

where the dots stand for terms that are shift invariant. Applying then $\bar D^4$ to this expression, one again obtains various terms. Although there is no $W^3 C$ term, the terms in $W^2 F^2$ are

\be   \varepsilon_{ijpqr} \varepsilon_{klstu} W^p W^s  \partial^2 F^{qr} \partial^2 F^{tu} \; , \ee

and 

\be  \varepsilon_{ijpqr} \varepsilon_{klstu} W^p \partial W^q  \partial F^{rs} \partial^2 F^{tu}
\; ,  \ee

both being projected into the $[0,0|0200]$. The term $W\chi\rho C$ also produces a term

\be \varepsilon_{ijpqr} W^p \chi \partial \chi_{kl} \partial^2 F^{qr} \; , \ee

projected into the $[0,0|0200]$. Once again, the shift variation of $M_{0,4,0}$  does not vanish, and is not a total derivative either. We therefore conclude that the unique $R^4$ invariant in $N=5$ supergravity is not $SU(5,1)$ invariant. 

To conclude this section, we have shown that duality invariance implies the absence of 3-loop divergences in $N=5, 6$ supergravity, and of 4-loop divergences in $N=6$ supergravity. In addition, as in the $N=8$ case, there are only harmonic measures of type $(1,1)$ in both of these theories. This means that there are non-linear measures for the $N=5$ $R^4$ and the $N=6$ $\del^2 R^4$ invariants, but there must be a violation of duality group symmetry as there are no duality invariant measures. On the other hand, there is no non-linear measure for  the $N=6$, $R^4$ invariant from which one would conclude that the corresponding non-linear cocycle is non-standard hence protected by algebraic renormalisation.


\section{Concluding remarks}


In this article, we have advanced field-theoretic arguments in favour of the idea that the short BPS invariants in $N=8$ supergravity fail to be $E_{7(7)}$ invariant. From this, one concludes that the onset of divergences should be postponed to at least seven loops, where there is a candidate $E_{7(7)}$ invariant counterterm, namely the volume of superspace. For the short invariants, we have presented arguments based on the impossibility of achieving a trivial scalar factor in front of the purely gravitational $R^4$, $\partial^4 R^4$ and $\partial^6 R^4$ terms because a non-trivial scalar factor is required by dimensional reduction and because the uniqueness of the linearised $D=4$ counterterms at the 3, 5 and 6 loop orders rules out the possibility of a cancellation between inequivalent terms coming from higher dimensions. We have also demonstrated that the $R^4$ invariant is indeed not $E_{7(7)}$ invariant by establishing the non-invariance of the last component of the associated linearised closed super four-form under constant shifts of the scalar fields, \ie under linearised $\ge_7$ transformations. This comes about because this linearised term will affect the four-gravitino term (an eight-point contribution) in the non-linear spacetime invariant.

In addition, we have investigated the question of whether appropriate measures exist in curved $N=8$ superspace.  In two cases, corresponding to the $R^4$ and $\partial^4 R^4$ invariants, the answer is no, whereas for the $\partial^6 R^4$ invariant a measure seems to be available. However, even in this case, there is no available integrand that could be $E_{7(7)}$ invariant as such an integrand would have to be constructed from the undifferentiated scalars. We stress that the non-existence of harmonic measures for the $R^4$ and $\partial^4 R^4$ invariants does not imply that there are no such invariants in the full theory. Indeed, our analysis of the $t_0$-cohomology in $N=8$ supergravity demonstrates that  in principle there exist closed super-four-forms whose structure is incompatible with the possibility of writing them as harmonic superspace integrals. This translates in components into the property that such invariants admit terms quartic in undifferentiated gravitino fields with a tensor structure that cannot appear in harmonic superspace integrals (at least without introducing a prepotential). This suggests that the non-linear $R^4$ and $\partial^4R^4$ invariants are associated to super-four-forms with a structure different from that of the other invariants, so that the supersymmetry Ward identities within the algebraic approach would by themselves be sufficient to rule out the possibility of the corresponding logarithmic divergences at 3 or 5-loops.

 A further aspect of this purely field-theoretic analysis is that there are UV divergence implications for supergravity theories with fewer supersymmetries. The $R^4$ counterterm is a BPS invariant for $N=5$ and $N=6$ and the superspace arguments given above adapt to these cases straightforwardly. Indeed, we have shown that the closed super four-forms associated to these counterterms are not invariant under constant shifts of the scalar fields, hence establishing that they are not duality-invariant. The same argument applies to the $ \partial^2 R^4 $ BPS invariant in $N=6$  (recalling that there is a linearised four-loop invariant in $N=6$, unlike the case of $N=8$).  It therefore follows that there are non-renormalisation theorems at three loops for $N=5,6$ and also at four loops for $N=6$. There are non-linear harmonic measures for $R^4$ in $N=5$ and $\del^2 R^4$ in $N=6$ but the corresponding integrands cannot be duality invariant, while the cocycle for $R^4$ in $N=6$ is non-standard thus providing additional evidence that these counterterms are protected by duality symmetries. The first divergences in these theories are therefore likely to occur at five and four loops for $N=6$ and $5$. The counterterm in both cases is the volume of superspace which should integrate to $\del^4 R^4$ for $N=6$ and $\del^2 R^4$ for $N=5$ (together with higher-order terms).
  
\section*{Note added}

After the first version of this article was posted to the arXiv, a paper discussing duality symmetries of invariants from a somewhat different perspective appeared \cite{Beisert:2010jx}.

\section*{Acknowledgements}

We thank Nicolas Beisert, Michael Green, Renata Kallosh, and Tristan McLoughlin for useful discussions. KSS would like to thank the Albert Einstein Institute, Potsdam for hospitality during the course of the work. The work of KSS was supported in part by by the STFC under rolling grant PP/D0744X/1.

\end{document}